\documentclass[10pt,USenglish,twocolumn]{article}

\usepackage[utf8]{inputenc}
\usepackage{lmodern}
\usepackage{algorithm}
\usepackage{algpseudocode}
\usepackage{amsmath,amssymb,amsthm}
\usepackage{MnSymbol}
\usepackage{graphicx}
\usepackage{xcolor}
\usepackage[numbers,square,sort&compress]{natbib}
\usepackage{subcaption}
\usepackage{units}
\usepackage[%
  colorlinks=true,%
  urlcolor=blue,%
  linkcolor=.,%
  citecolor=.,%
  pageanchor=false,%
]{hyperref}
\usepackage[left=2cm,right=2cm,top=2cm,bottom=2cm]{geometry}
\usepackage{titlesec}
\titleformat{\title}
  {\normalfont\sffamily\Huge\bfseries}
  {\thetitle}{1em}{}
\titleformat{\section}
  {\normalfont\sffamily\Large\bfseries}
  {\thesection}{1em}{}
\titleformat{\subsection}
  {\normalfont\sffamily\large\bfseries}
  {\thesubsection}{1em}{}

\theoremstyle{definition}
\newtheorem{alg}{Algorithm}

\newcommand{\Paragraph}[1]{\paragraph{\textsf{#1}}}

\makeatletter
\renewcommand*\env@matrix[1][\arraystretch]{%
  \edef\arraystretch{#1}%
  \hskip -\arraycolsep
  \let\@ifnextchar\new@ifnextchar
  \array{*\c@MaxMatrixCols c}}
\makeatother

\newcommand\infoFootnote[1]{%
  \begingroup
  \renewcommand\thefootnote{}\footnote{#1}%
  \addtocounter{footnote}{-1}%
  \endgroup}

\newcommand{\R}{\mathbb{R}}
\newcommand{\Z}{\mathbb{Z}}
\newcommand{\N}{\mathbb{N}}

\newcommand{\Kc}{\mathcal{K}}
\newcommand{\Ic}{\mathcal{I}}
\newcommand{\Nc}{\mathcal{N}}
\newcommand{\Mc}{\mathcal{M}}
\newcommand{\Vc}{\mathcal{V}}

\newcommand{\Ac}{\mathcal{A}}
\newcommand{\Tc}{\mathcal{T}}
\newcommand{\Oc}{\mathcal{O}}

\newcommand{\sk}{\mathtt{sk}}
\newcommand{\pk}{\mathtt{pk}}
\newcommand{\Enc}{\mathsf{Enc}}
\newcommand{\Dec}{\mathsf{Dec}}

\newcommand{\roundlr}[1]{\left\lfloor {#1}\right\rceil}

\newcommand{\cipher}[1]{\lsem {#1}\rsem}

\begin{document}

\title{{\normalfont\sffamily\huge\bfseries Towards privacy-preserving cooperative control via encrypted distributed optimization}}
\author{Philipp Binfet\thanks{The first three authors contributed equally to this work.}
\and Janis Adamek\footnotemark[\value{footnote}]
\and Nils Schlüter\footnotemark[\value{footnote}]
\and Moritz Schulze Darup\vspace{2mm}}
\date{}

\maketitle

\Paragraph{Abstract:}
Cooperative control is crucial for the effective operation of dynamical multi-agent systems.
Especially for distributed control schemes, it is essential to exchange data between the agents.
This becomes a privacy threat if the data is sensitive.
Encrypted control has shown the potential to address this risk and ensure confidentiality. However, existing approaches mainly focus on cloud-based control and distributed schemes are restrictive.

In this paper, we present a novel privacy-preserving cooperative control scheme based on encrypted distributed optimization. More precisely, we focus on a secure distributed solution of a general consensus problem, which has manifold applications in cooperative control, by means of the alternating direction method of multipliers (ADMM).
As a unique feature of our approach, we explicitly take into account the common situation that local decision variables contain copies of quantities associated with neighboring agents and ensure the neighbor's privacy. We show the effectiveness of our method based on a numerical case study dealing with the formation of mobile robots.
\infoFootnote{This is an Accepted Manuscript of an article published by De Gruyter in {at~-~Automatisierungstechnik}, available at http://www.degruyter.com/ \href{https://doi.org/10.1515/auto-2023-0082}{https://doi.org/10.1515/auto-2023-0082}}
\infoFootnote{P.~Binfet, J.~Adamek, N.~Schlüter, M.~Schulze Darup are with the Control and Cyberphysical Systems Group at TU Dortmund University, e-mail: \href{mailto:moritz.schulzdarup@tu-dortmund.de}{moritz.schulzdarup@tu-dortmund.de}}

\Paragraph{Keywords:}
Privacy-preserving control,
cooperative control,
distributed optimization,
homomorphic encryption,
multi-agent systems,
cyber-physical systems

\section{Introduction}
Recent advances in communication technology and cloud computing lead to an increased connectivity of control systems. The interconnection of these (cyber-physical) systems enables new features and can offer
improved efficiency, performance, and safety. Examples range from vehicle to vehicle communication for collision avoidance in road traffic to process optimization of spatially distributed factories.
However, sensitive data can hinder these applications because external processing on not necessarily trustworthy platforms is required.
In principle, tailored cryptosystems such as homomorphic encryption (HE, see \cite{marcolla2022survey} for an illustrative survey) can solve this issue as they allow for privacy-preserving computations on encrypted data. However, the application of these cryptosytems especially for control purposes is non-trivial and has motivated the young research field of encrypted control (see \cite{SchulzeDarup2021_CSM} for an overview). Tailored encrypted implementations have been realized for various control schemes, including linear feedback \cite{Kogiso2015,Farokhi2017,Kim2022_TAC} and model predictive control \cite{SchulzeDarup2018_LCSS,Alexandru2018_CDC}. However, most of the existing schemes consider cloud-based centralized control. In fact, encrypted distributed control for multi-agent systems (MAS) is still in its infancy.

In this paper, we focus on privacy-preserving cooperative control by means of distributed optimization.
Distributed optimal control is essential for various applications such as electrical power grids \cite{maneesha2021powergridsurvey} or robot formation \cite{kwang2015surveymultiagentformationcontrol,van2017distributed}. Privacy-preserving implementations offer promising features, e.g., for the usage of smart meters or for autonomous driving. In order to leverage this potential, flexible and universally usable implementations are required.
Hence, we use the general consensus problem and a solution via the alternating direction method of multipliers (ADMM, see \cite{boyd2011admm}) as a basis. Remarkably, privacy-preserving general consensus and distributed ADMM have both been addressed before, and we briefly summarize related work next. However, we argue that crucial features for an extensive use have not been taken into account yet, which motivates our contribution.

Summarizing the state of the art, we first note that not only HE is an enabler for privacy-preserving computations.
In fact, secure multi-party computation with secret sharing \cite{lindell2020secure} and differential privacy  \cite{hassan2019differential} have also been used for privacy-preserving distributed optimization methods \cite{Tian2022PPpowersystems,tjell2019PPDistOpt,nozari2016differentially}.
However, differential privacy typically builds on adding noise to the data and multi-party computation often requires a tremendous amount of communication between the involved parties. Both drawbacks can be avoided with HE, and we consequently focus on this technology.
Moreover, while the mentioned existing schemes provide privacy for the local cost functions or decision variables, they neglect that the local variables often contain copies of quantities related to neighboring agents.
In contrast, our approach explicitly takes this condition into account and allows using such copies without leaking information about the neighboring agents. A similar goal is underlying the encrypted ADMM scheme in~\cite{Zhang2019admmDecentralizedOpt}. However, there, privacy is only provided for iterates of the algorithm and information is leaked after convergence, whereas we can ensure privacy also after convergence.
Further, we allow the local cost functions, which often contain private information of, e.g., the agent's dynamics, to be securely parameterized by a system operator in order to specify cooperative control goals.

We will illustrate our approach based on a case study on robot formation. In this context, we briefly note that encrypted control for mobile robots has been previously addressed in \cite{SchulzeDarup2019_LCSS,Alexandru2019_CDC,Marcantoni2023_LCSS}.
However, the two former contributions deal with distributed privacy-preserving state feedback control, which comes with limited application possibilities.
The latter uses a central computing unit and only distributed sensing and acting.

The remaining paper is organized as follows. In Section~\ref{sec:problemStatement}, we state the problem of interest and summarize the security goals.
Section~\ref{sec:distributedADMM} deals with the plaintext ADMM algorithm that solves our control problem in a distributed fashion. Our main contribution is stated in Section \ref{sec:encryptedImplementation}, where we develop our encrypted ADMM realization.
The case study on robot formation control is presented in Section~\ref{sec:robotFormation}. Finally, we conclude our findings and discuss future research directions in Section \ref{sec:ConclusionOutlook}.

\subsection*{Notation}

We mainly use standard notation such as $\R$, $\Z$, and $\N$ for the set of reals, integers, and natural numbers (including~$0$), respectively. A slightly special notion is devoted to the MAS setup and the ADMM iterations to be analyzed. In fact, the subscript of a variable such as $i$ in $z_i$ will refer to agent $i$. In contrast, the superscript as, e.g., $\tau$ in $z^{\tau}_{i}$ will refer to iterations (or blocks of matrices). To address the $k$-th entry of vector-valued variables, we use brackets around that variable with an index $k$ such as $(z_i)_k$. The index can also take the form of an (ordered) index set and then address multiple entries at once.
Special notation for encryption-related operations will be introduced in Section~\ref{subsec:essentialsHomEncr}.
\section{Problem statement}%
\label{sec:problemStatement}

\begin{figure*}[tp]
    \centering

    \begin{subfigure}[c]{0.325\textwidth}
    \captionsetup{justification=centering}
    \centering
    \subcaption{Ring graph}%
    \label{subfig:ring}
    \vspace{1mm}
    \includegraphics[width=0.7\textwidth]{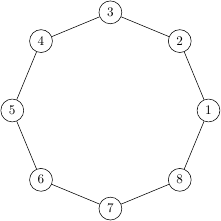}
    \end{subfigure}
    \begin{subfigure}[c]{0.325\textwidth}
    \captionsetup{justification=centering}
    \centering
    \subcaption{Star graph}%
    \label{subfig:star}
    \vspace{1mm}
    \includegraphics[width=0.7\textwidth]{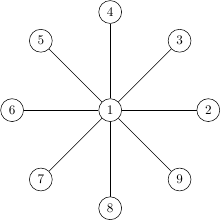}
    \end{subfigure}
    \begin{subfigure}[c]{0.325\textwidth}
    \captionsetup{justification=centering}
    \centering
    \subcaption{Generic graph}%
    \label{subfig:generic}
    \vspace{1mm}
    \includegraphics[width=0.7\textwidth]{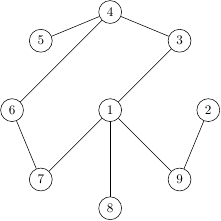}
    \end{subfigure}

     \caption{Exemplary communication graphs. The first two graphs offer specific structures, while the latter is generic.}%
    \label{fig:threeGraphs}
\end{figure*}

We consider a multi-agent system (MAS) with $M$ dynamical agents. The agents can exchange information via a bidirectional communication network described by an undirected graph (see, e.g., Fig.~\ref{fig:threeGraphs}). In addition, a system operator can communicate with each agent.
The operator specifies the cooperative control task by parametrizing a (separable) cost function to be minimized by the agents. More precisely, we assume that the agents are supposed to (approximately) solve a general consensus problem of the form (cf.~\cite[Sect.~7.2]{boyd2011admm})
\begin{equation}
\label{eq:generalizedConsensusOP}
    \min_{z_1,\dots,z_M,\zeta} \,\sum_{i=1}^M f_i(z_i,p_i) \qquad \text{s.t.} \qquad z_i =  (\zeta)_{\Kc_i}
\end{equation}
at each sampling instance, where the constraints apply for each $i \in \Mc:=\{1,\dots,M\}$. Here, $z_i \in \R^{v_i}$, $p_i\in\R^{w_i}$, and $f_i:\R^{v_i} \times \R^{w_i} \rightarrow \R$
denote local decision variables, parameters, and cost functions,
respectively, and ${\zeta \in \R^\nu}$ refers to a global decision variable.
Furthermore, the (ordered) index sets $\Kc_i\subseteq\Vc:=\{1,\dots,\nu\}$
specify which entries of the global $\zeta$ are associated with the local~$z_i$. Regarding the cost functions, we assume that each consist of two terms: A quadratic form in $z_i$ with linear parameter dependencies and an indicator function-like term reflecting affine equality constraints. More formally, we consider
\begin{equation}
\label{eq:fiQuadratic}
    f_i(z_i,p_i):=\frac{1}{2} z_i^\top H_i z_i + p_i^\top F_i^\top z_i+ \left\{ \begin{array}{ll}
      \!0   &  \!\!\text{if} \,\, G_i z_i = E_i p_i,\!\!\\
      \!\infty   &  \!\!\text{otherwise},
    \end{array}\right.
\end{equation}
where the equality constraints will allow incorporating the dynamics of the agents.

Now, we aim for a distributed but privacy-preserving solution of~\eqref{eq:generalizedConsensusOP}. To specify the corresponding security goals, we make the following non-standard assumptions which, however, reflect common setups. We assume that $z_i$ and $p_i$ offer the  structures
\begin{equation}
    \label{eq:zipiStructures}
z_i:=\begin{pmatrix}
\alpha_{i} \\
\gamma_{i}
\end{pmatrix} \qquad \text{respectively} \qquad p_i:=\begin{pmatrix}
\beta_{i} \\
\delta_{i}
\end{pmatrix}
\end{equation}
(with potentially different block dimensions), where $\alpha_i$ and $\beta_i$ reflect individual quantities of agent $i$ (such as individual inputs, states, or outputs). Further, $\gamma_i$ stands for quantities associated with neighboring agents and $\delta_i$ reflects parameters set by the operator (e.g., a parameterization of a formation).
Next, taking into account the set of neighbors $\Nc_i \subseteq \Mc\setminus\{i\}$ of agent~$i$, allows us to specify the role of $\gamma_i$.
In fact, we assume that the $\gamma_i$ represent selected entries of the individual quantities $\alpha_j$ of neighboring agents $j \in \Nc_i$.
As a consequence and due to the constraints $z_i =  (\zeta)_{\Kc_i}$,
solving~\eqref{eq:generalizedConsensusOP} results in
\begin{equation}
\label{eq:zeta}
  \zeta^\ast = \begin{pmatrix}
    \alpha_1^\ast\\
    \vdots \\
    \alpha_M^\ast
\end{pmatrix}
\end{equation}
for suitable sorting and choices of $\Kc_i$.
We are now ready to specify the security goals as follows:

\vspace{2mm}
\begin{enumerate}
    \item Only agent $i$ should have access to plaintext values of $\alpha_i$ and $\beta_i$
    as well as $H_i$,~$F_i$, $G_i$, and $E_i$.
    \item Only the operator should have access to plaintext values of $\delta_i$.
    \item Neither the agents nor the operator should have access to plaintext values of any $\gamma_i$.
    \item Neither the agents nor the operator should have access to plaintext values of $\zeta$, with the exception that agent~$i$ may have access to those entries reflecting $\alpha_i$.
\end{enumerate}
In other words, while solving~\eqref{eq:generalizedConsensusOP} in a distributed manner, the agents should not gain any local or global insights apart from their individual quantities. Further, (if not desired for an application) the operator should not get any insights into the obtained solution to~\eqref{eq:generalizedConsensusOP} but should be able to specify the cooperative control task by means of the $\delta_i$. Achieving these security goals also requires specifying an attacker model. In this work, we assume that the agents are honest-but-curious (semi-honest), i.e., they will faithfully execute computations and communications prescribed by the protocol but may otherwise use all information available to them to gain knowledge about secret data.
We briefly note that the specified security goals are more strict than in the existing literature. For instance, in \cite{Zhang2019admmDecentralizedOpt}, neighboring agents get access to $\alpha_i$ after convergence.

\section{Distributed solution via ADMM}%
\label{sec:distributedADMM}

Before focussing on a privacy-preserving implementation in Section~\ref{sec:encryptedImplementation}, we briefly present a distributed (plaintext) solution of~\eqref{eq:generalizedConsensusOP} via ADMM. The largely standard procedure builds on the augmented Lagrangian
\begin{align*}
&L_\rho(z_1,\dots,z_M,\zeta,\lambda_1,\dots,\lambda_M):=\\
&\quad \sum_{i=1}^M f_i(z_i,p_i) + \lambda_i^\top (z_i - (\zeta)_{\Kc_i}) + \frac{\rho}{2} \| z_i - (\zeta)_{\Kc_i} \|_2^2
\end{align*}
with the dual variables $\lambda_i \in \R^{v_i}$ and the positive weighting factor $\rho\in \R$.
According to \cite[p.~55]{boyd2011admm}, the three ADMM steps per iteration $\tau \in \N$ are
\begin{subequations}%
\label{eq:stepsADMM}
\begin{align}
    \label{eq:firstStepADMM}
    z_i^{\tau+1}&:=\arg \min_{z_i} \, L_\rho(z_1,\dots,z_M,\zeta^\tau,\lambda_1^\tau,\dots,\lambda_M^\tau), \\
    \label{eq:secondStepADMM}
    \zeta^{\tau+1}&:=\arg \min_{\zeta} \, L_\rho(z_1^{\tau+1},\dots,z_M^{\tau+1},\zeta,\lambda_1^\tau,\dots,\lambda_M^\tau), \\
    \label{eq:lambdaUpdate}
    \lambda_i^{\tau+1} &:=\lambda_i^{\tau} +\rho \left( z_i^{\tau+1} - (\zeta^{\tau+1})_{\Kc_i} \right).
\end{align}
\end{subequations}

\subsection{Specific ADMM iterations}

For the specified problem~\eqref{eq:generalizedConsensusOP} with the cost functions~\eqref{eq:fiQuadratic} the update $z_i^{\tau+1}$ follows from the solution of the equality-constrained quadratic program
\begin{align*}
    &\min_{z_i} \, \frac{1}{2} z_i^\top H_i z_i + p_i^\top F_i^\top z_i + \lambda_i^{\tau,\top} z_i  + \frac{\rho}{2} \| z_i - (\zeta^{\tau})_{\Kc_i}  \|_2^2 \\
&\,\,\,\text{s.t.} \quad G_i z_i = E_i p_i
\end{align*}
and we consequently find $z_i^{\tau+1}$ from solving
\begin{equation}
\label{eq:ziUpdate}
    \begin{pmatrix}
        H_i + \rho I_{v_i}  & G_i^\top \\
        G_i & 0
    \end{pmatrix}
    \begin{pmatrix}
         z_i^{\tau+1} \\
         \mu_i^{\tau+1}
    \end{pmatrix} = \begin{pmatrix}
         \rho  (\zeta^{\tau})_{\Kc_i}  - F_i p_i -\lambda_i^\tau  \\
         E_i p_i
    \end{pmatrix},
\end{equation}
where the Lagrange multipliers $\mu_i^{\tau+1}$ associated with the equality constraints are irrelevant here.
Regarding the second ADMM step~\eqref{eq:secondStepADMM},  we find
\begin{equation}
\label{eq:zetaUpdateLocalAvg}
 \left(\zeta^{\tau+1}\right)_k=\frac{1}{|\Ic_k|} \sum_{i\in \Ic_k} (z_i^{\tau+1})_{k_i}
\end{equation}
under the assumption that $\lambda_i^0:=0$,
where $k_i$ refers to the local position of the global entry $k$ and where the sets
\begin{equation}
\nonumber
  \Ic_k:=\{ i \in \Mc \,|\, k \in \Kc_i \}
\end{equation}
collect all agents $i$ who make use of
$(\zeta)_k$.
In this context, we assume that, for each $k \in \Vc$, there exists an agent $i\in \Mc$ such that
\begin{equation}
\label{eq:IkSubseteqNi}
\Ic_k \subseteq \Nc_i \cup \{i\}.
\end{equation}
In other words, for each entry of the global variable, there exists at least one agent $i$ who is able to evaluate~\eqref{eq:zetaUpdateLocalAvg} by collecting information only from its neighbors $j\in \Nc_i$ and performing local averaging.
Remarkably, \eqref{eq:IkSubseteqNi} is in line with the assumption that $\gamma_i$ reflects quantities in $\{\alpha_j \,|\, j \in \Nc_i\}$.
In fact, a violation of \eqref{eq:IkSubseteqNi} implies that there exists a $\gamma_i$ which contains elements of some $\alpha_j$ with $j\notin \Nc_i$.

\subsection{Job scheduling}

The distributed solution of~\eqref{eq:generalizedConsensusOP} via ADMM can be carried out as follows. Each agent starts with $\lambda_i^0:=0$ and some initial guess for $(\zeta^0)_{\Kc_i}$ (possibly obtained by collecting data from neighbors) and computes $z_i^1$ via \eqref{eq:ziUpdate}. Next, for each $k\in \Vc$, the local averaging \eqref{eq:zetaUpdateLocalAvg} is performed by one agent $i$ satisfying \eqref{eq:IkSubseteqNi}.
To this end, the corresponding agent collects $(z_j^{1})_{k_j}$ from each of its neighbors $j \in \Ic_k \setminus \{i\} \subseteq \Nc_i$ and returns $\left(\zeta^{1}\right)_k$ to these neighbors after performing the averaging.
At this point, all agents hold the required data to compute their
$\lambda_i^1$ according to~\eqref{eq:lambdaUpdate}.
This completes the first ADMM iteration, and all following ones are carried out analogously. Under the assumption that all $H_i$ are positive definite, we have convexity of the $f_i$ and, consequently, the ADMM iterations are guaranteed to converge to an optimum \cite[Sect.~3.2.1]{boyd2011admm}.

Now, given the layout~\eqref{eq:zeta} of the global variable, we can specify the local averaging \eqref{eq:zetaUpdateLocalAvg}. In fact, a generic setup then is to let agent~$i$ perform the averaging for all entries~$k$ of $\zeta$ associated with~$\alpha_i$.
We collect all these entries in the index set $\Ac_i \subseteq \Vc$ and find, by construction,
$$
\Ac_i \cap \Kc_j \neq \emptyset  \quad \implies \quad j \in \Nc_i
$$
for all $(i,j) \in \Mc^2$ with $i\neq j$. During iteration step $\tau$, agent $i$ then evaluates the following algorithm, where we emphasize the communication with neighbors as a preparation for the encrypted implementation in Section~\ref{sec:encryptedImplementation}.
\begin{alg}%
\label{alg:plainADMM}
ADMM iteration $\tau$ at agent $i$.

\begin{enumerate}
    \item Compute $z_i^{\tau+1}$ via \eqref{eq:ziUpdate} for given $(\zeta^\tau)_{\Kc_i}$, $\lambda^\tau_i$, and $p_i$.
    \item For each $k \in \Kc_i \setminus \Ac_i$ and each $j \in \Nc_i$, if $k \in \Ac_j$, send $(z_i^{\tau+1})_{k_i}$ to agent $j$.
    \item For each $k \in \Ac_i$ and each $j \in \Nc_i$, if $k \in \Kc_j$, receive $(z_j^{\tau+1})_{k_j}$ from agent $j$.
    \item Compute $(\zeta^{\tau+1})_{\Ac_i}$ according to~\eqref{eq:zetaUpdateLocalAvg}.
    \item For each $j\in \Nc_i$, send $(\zeta^{\tau+1})_{\Ac_i \cap \Kc_j}$ to agent $j$.
    \item For each $j\in \Nc_i$, receive $(\zeta^{\tau+1})_{\Ac_j \cap \Kc_i}$ from agent $j$.
    \item Compute $\lambda_i^{\tau+1}$ according to~\eqref{eq:lambdaUpdate}.
\end{enumerate}
\end{alg}
Clearly, the computations in
steps
1, 4, and~7 refer to the three ADMM steps. In contrast,
steps
2, 3, 5, and~6 reflect data exchange required for performing the local averaging and distributing the corresponding results. We complete our tailored ADMM implementation by noting that the layout~\eqref{eq:zeta} also promotes the following initialization.

\begin{alg}%
\label{alg:plainInit}
ADMM initialization at agent $i$ based on a guess  $\hat{\alpha}_i$ for $\alpha_i^\ast$.
    \begin{enumerate}
    \item Set $\lambda_i^0:=0$ and, for each $k \in \Ac_i$, set $(\zeta^0)_{k}:=(\hat{\alpha}_i)_{k_i}$.
    \item For each $k \in \Ac_i$ and each $j \in \Nc_i$, if $k \in \Kc_j\setminus \Ac_j$, send $(\hat{\alpha}_i)_{k_i}$ to agent $j$.
    \item For each $k \in \Kc_i \setminus \Ac_i$, receive $(\hat{\alpha}_j)_{k_j}$ from the agent $j$ for which $k\in \Ac_j$, and set $(\zeta^0)_{k}:=(\hat{\alpha}_j)_{k_j}$.
\end{enumerate}
\end{alg}

\section{Encrypted implementation}%
\label{sec:encryptedImplementation}

In the following, we will use HE to implement the ADMM-based solution to~\eqref{eq:generalizedConsensusOP} or, more specifically, Algorithms~\ref{alg:plainADMM} and \ref{alg:plainInit}
in a privacy-preserving manner according to the security goals in Section~\ref{sec:problemStatement}. It will turn out that the structures of the algorithms can be preserved. However, securely evaluating the updates $z_i^{\tau+1}$, $(\zeta^{\tau+1})_{\Ac_i}$, and $\lambda_i^{\tau+1}$ at agent~$i$ requires encrypting these computations using ciphers, which are not decryptable by agent $i$. As specified below, we solve this issue by utilizing multiple instances of a homomorphic cryptosystem.
More precisely, each agent will set up its own instance $i$. In addition, the operator sets up the instance $0$.
We will then use the operator's instance, with ciphers that cannot be decrypted by the agents, to securely implement the ADMM scheme.
However, after evaluating $\ell\in \N$ encrypted iterations, agent $i$ shall obtain the entries associated with $\alpha_i$ in $z_i^\ell$, which we denote as~$\alpha_i^\ell$ from now on. In order to realize this,
we need to be able to transform ciphers from one of the cryptosystem's instances to another. We next summarize the corresponding technique of key switches together with other essential features of HE, before presenting the encrypted implementation of Algorithms~\ref{alg:plainADMM} and \ref{alg:plainInit} in Section~\ref{subsec:encryptedADMMviaKeySwitches}.

\subsection{Essentials on homomorphic encryption}%
\label{subsec:essentialsHomEncr}
HE enables computations on ciphertexts such that the privacy of the underlying plaintexts is preserved \cite{marcolla2022survey}.
More formally, given two integer numbers $x$ and $y$,
a homomorphic cryptosystems can provide the basic operations ``$\oplus$'' and ``$\otimes$'' such that the relations
\begin{subequations}%
\label{eq:homomorphisms}
\begin{align}%
\label{eq:addHomomorphism}
    \Dec\left(\Enc(x) \oplus \Enc(y) \right) &= x+y, \\
    \label{eq:mulHomomorphism}
    \Dec\left(\Enc(x) \otimes \Enc(y) \right) &= x y
\end{align}
\end{subequations}
hold, where $\Enc$ and $\Dec$ refer to the encryption and decryption procedure. In other words, \eqref{eq:addHomomorphism} and \eqref{eq:mulHomomorphism} allow to securely carry out additions and multiplications, respectively, based on the ciphertexts $\Enc(x)$ and $\Enc(y)$.
More specifically, we call a cryptosystem additively homomorphic if \eqref{eq:addHomomorphism} is supported and multiplicatively homomorphic if \eqref{eq:mulHomomorphism} is given.
Fully homomorphic encryption (FHE) enables both encrypted operations arbitrarily often, whereas a cryptosystem that provides a limited amount of consecutive encrypted operations (especially multiplications)
is referred to as leveled FHE.
Remarkably, FHE typically builds on a leveled FHE scheme, which is extended by a computationally demanding routine called bootstrapping. Loosely speaking, this routine allows to
refresh a ciphertext after the computation limit is reached.
Now, it will turn out that additive HE is, in principle, sufficient to encrypt the ADMM iterations. However, securely exchanging information between the agents will require a procedure called key switching, which is not provided by common additively homomorphic schemes such as \cite{Paillier1999}. Hence, we will later use leveled FHE but avoid  bootstrapping for efficiency reasons.
Implementation-wise, leveled FHE schemes
usually built on
variants of the learning with errors (LWE) problem~\cite{regev2009lattices}. Explaining details of the corresponding cryptosystems is beyond the scope of this paper. However, we briefly summarize relevant properties
for our implementation.

First, any (at least) additively homomorphic cryptosystem also supports partially encrypted multiplications. In fact, for integers $x$ and $y$, we obviously have
\begin{equation}
\label{eq:mulConstHomomorphism}
x\,\Enc(y) = \underbrace{\Enc(y) \oplus \dots \oplus \Enc(y)}_{x\,\text{times}}=x \odot \Enc(y),
\end{equation}
where we note that many cryptosystems (including LWE-based schemes) provide a dedicated and efficient operation ``$\odot$'', which does not rely on multiple additions.
Furthermore, additively homomorphic cryptosystem also support encrypted subtractions, which we denote by ``$\ominus$''.
Second, the cryptosystems of interest consider messages $x$ from a finite set such as, e.g.,
the set of integers modulo $q$ denoted by $\Z_q$.
Staying within this set during the encrypted operations, i.e., avoiding overflows, is crucial for flawless operation.
Third, we use public key schemes, i.e., encryptions are enabled
by a public key $\pk$ and only decryptions require the secret $\sk$. Hence, encryptions and also homomorphic operations can be carried out by
any party, but only a party holding the secret key is able to recover the corresponding plaintexts.
Fourth, we can consider multiple
instances
of a cryptosystem and use key switches
to transfer ciphertexts from one instance
to another.
We denote the public and secret key of an instance $i$ with $\pk_i$ and $\sk_i$, respectively.
Further, $\Enc_i$ and $\Dec_i$ refer to the corresponding encryption and decryption procedures, respectively. Finally,
we can transform ciphertexts $\Enc_i(x)$ to $\Enc_j(x)$ by homomorphically switching the underlying secret key from $\sk_i$ to $\sk_j$ (see \cite[App.~B]{kim2021revisiting} for an overview). Such procedures, which we will abbreviate with $\Enc_{i\rightarrow j}(x)$, require $\Enc_j(\sk_i)$, i.e., an encryption of the secret key $\sk_i$
by means of $\Enc_j$,
and can then efficiently be realized based on homomorphic addition
and ``$\odot$'' (see
\cite[Sect.~4.2]{cheon2018bootstrapping} for details).

Now, applying a leveled FHE scheme to encrypt computations requires some preparation. First, messages to be encrypted need to be integers within the supported finite set. For instance, evaluating the updates~\eqref{eq:zetaUpdateLocalAvg} in an encrypted fashion requires mapping the $(z_i^{\tau+1})_{k_i}$ to, e.g., $\Z_q$. We use a simple but established mapping
\begin{equation}
\label{eq:scalingRoundingModq}
 \roundlr{%
    s \cdot (z_i^{\tau+1})_{k_i}
 } \!\mod q,
\end{equation}
which builds on a scaling by some positive factor $s \geq 1$, followed by rounding to the nearest integer and reducing the result modulo~$q$. The resulting integer can then be encrypted using, e.g., the encryption procedure $\Enc_0$ of the operator.
In order to simplify the notation and to make the link to the original data more explicit, we will use the shorthand notation
$\cipher{(z_i^{\tau+1})_{k_i}}_0$
for the composition of the mapping~\eqref{eq:scalingRoundingModq} and the encryption $\Enc_0$. We will also extend the notation to the homomorphic operations and, e.g., consider
\begin{equation}
\label{eq:encryptedZetaUpdate}
\cipher{(\zeta^{\tau+1})_{k} }_0 = \frac{1}{|\Ic_k|} \odot \left(\bigoplus_{i \in \Ic_k} \cipher{ (z_i^{\tau+1})_{k_i}}_0\right)
\end{equation}
as an encrypted version of~\eqref{eq:zetaUpdateLocalAvg}, where ``$\bigoplus_{i}$'' stands for sums over $i$ evaluated with ``$\oplus$'' and where we do not write the mapping via~\eqref{eq:scalingRoundingModq} for plaintext constants (as $1/|\Ic_k|$) explicitly.
Further, we extend the notation to vector-valued and matrix-valued arguments and, e.g., consider
\begin{equation}
\label{eq:encrpytedLambdaUpdate}
\cipher{\lambda_i^{\tau+1} }_0 = \cipher{\lambda_i^{\tau} }_0 \oplus \left( \rho \odot \left( \lsem z_i^{\tau+1} \rsem_0 \ominus \lsem (\zeta^{\tau+1})_{\Kc_i} \rsem_0 \right) \right)
    \end{equation}
as an encrypted version of~\eqref{eq:lambdaUpdate}.
Finally, inspired by $\Enc_{i\rightarrow j}$, we denote key switches from, e.g., $\cipher{\alpha_i^\ell}_{0}$ to $\cipher{\alpha_i^\ell}_{i}$ with $\cipher{\alpha_i^\ell}_{0\rightarrow i}$.
It remains to comment on the value $q$
by which
the finite set
and
the mapping~\eqref{eq:scalingRoundingModq}
are parameterized.
Roughly speaking, the supported level (i.e., the number of consecutive encrypted multiplications)
but also the computational effort (of encryptions, homomorphic operations, and decryptions) increase with $q$ while the risk of overflow decreases with $q$.
Hence, $q$ should be chosen large enough to support sufficiently many consecutive operations and to avoid overflows but otherwise as small as possible for computational efficiency.

\subsection{Encrypted ADMM via key switches}%
\label{subsec:encryptedADMMviaKeySwitches}

We already prepared the encrypted implementation of Algorithm~\ref{alg:plainADMM} by means of~\eqref{eq:encryptedZetaUpdate} and \eqref{eq:encrpytedLambdaUpdate}.
It remains to specify the encrypted update of $z_i$ in the first step of Algorithm~\ref{alg:plainADMM} and secure data exchanges between the agents.
In plaintext, updating $z_i$ is carried out by solving~\eqref{eq:ziUpdate} for $z_i^{\tau+1}$.
More formally, given the (block) inverse
\begin{equation}
\label{eq:GammaiBlocks}
\begin{pmatrix}
    \Gamma_i^{11} & \Gamma_i^{12} \\
    \ast & \ast
\end{pmatrix} :=\begin{pmatrix}
    H_i + \rho I_{v_i}  & G_i^\top \\
    G_i & 0
\end{pmatrix}^{-1}\!,
\end{equation}
we obtain
\begin{align}
\nonumber
z_i^{\tau+1} & = \Gamma_i^{11} \left(  \rho  (\zeta^{\tau})_{\Kc_i}  - F_i p_i -\lambda_i^\tau  \right) + \Gamma_i^{12} E_i p_i \\
\label{eq:ziUpdateExplicit}
& = \rho  \Gamma_i^{11} (\zeta^{\tau})_{\Kc_i} - \Gamma_i^{11}  \lambda_i^\tau + \left( \Gamma_i^{12} E_i - \Gamma_i^{11} F_i  \right) p_i.
\end{align}
Now, based on~\eqref{eq:ziUpdateExplicit}, it is tempting to realize the encrypted update of $z_i$ via
\begin{align}
\nonumber
   \lsem z_i^{\tau+1} \rsem_{0} &= \left( \rho \Gamma_i^{11}\odot \lsem (\zeta^\tau)_{\Kc_i} \rsem_{0} \right) \ominus \left( \Gamma_i^{11} \odot\lsem \lambda_i^\tau \rsem_{0} \right) \\
   \label{eq:encryptedZiUpdate}
   &\quad \oplus  \left( (\Gamma_i^{12} E_{i} -\Gamma_i^{11} F_i ) \odot  \lsem p_i \rsem_{0} \right).
\end{align}
However, in order to actually use this realization, we have to specify how the various parameters and variables are securely provided.
Under the assumption that the ADMM parameter $\rho$ is known by the agents, the blocks $\Gamma_i^{11}$ and $\Gamma_i^{12}$ in~\eqref{eq:GammaiBlocks} can be precomputed by agent $i$, who also has access to the required matrices $H_i$ and $G_i$ according to the security goals in Section~\ref{sec:problemStatement}.
Further, agent $i$ can precompute the deduced quantities $\rho \Gamma_i^{11}$
and $\Gamma_i^{12} E_{i} -\Gamma_i^{11} F_i$.
Regarding the parameter $p_i$, the block $\beta_i$ can likewise be provided and encrypted by agent $i$. In contrast, $\cipher{\delta_i}_0$ has to be delivered by the operator.
The remaining variables $(\zeta^\tau)_{\Kc_i}$ and $\lambda_i^\tau$ are assumed to be given in Algorithm~\ref{alg:plainADMM} since they are either initialized according to Algorithm~\ref{alg:plainInit} or computed during the previous ADMM iteration $\tau-1$.
We adopt this setup for the encrypted implementation, and will ensure
that $\cipher{(\zeta^\tau)_{\Kc_i} }_{0}$ and $\cipher{\lambda_i^\tau}_{0}$ are available when evaluating \eqref{eq:encryptedZiUpdate}.

Given the encrypted versions \eqref{eq:encryptedZetaUpdate}, \eqref{eq:encrpytedLambdaUpdate}, and \eqref{eq:encryptedZiUpdate} of the three ADMM steps, we are almost ready to encrypt Algorithm~\ref{alg:plainADMM}.
However, sharing elements of $\cipher{z_i^{\tau+1} }_{0}$ and $\cipher{(\zeta^{\tau+1})_{\Ac_i} }_{0}$ with neighbors analogously to steps 2 and 5 of Algorithm~\ref{alg:plainADMM} shall not be carried out without another security layer. In fact, since the operator holds $\sk_0$ and might be able to eavesdrop communications, there is a risk of leaking information about $z_i^{\tau+1}$ or $\zeta^{\tau+1}$, which violates the security goals. Fortunately, this issue can be easily solved by additionally encrypting the communication between the agents using established and efficient symmetric encryption schemes such as AES (i.e., the Advanced Encryption Standard).
We only briefly mention the additional (onion) encryption in the following encrypted version of Algorithm~\ref{alg:plainADMM} in order to keep the focus on the encrypted computations.

\begin{alg}%
\label{alg:encryptedADMM}
Encrypted ADMM iteration $\tau$ at agent $i$.

\begin{enumerate}
    \item Compute $\cipher{ z_i^{\tau+1} }_{0}$ according to~\eqref{eq:encryptedZiUpdate}  for given $\lsem (\zeta^\tau)_{\Kc_i} \rsem_{0}$, $\lsem \lambda^\tau_i \rsem_{0}$, and $\lsem p_i \rsem_{0}$.
    \item For each $k \in \Kc_i \setminus \Ac_i$ and each $j \in \Nc_i$, if $k \in \Ac_j$,   send $\cipher{(z_i^{\tau+1})_{k_i} }_{0}$ to agent $j$ secured by AES.
    \item For each $k \in \Ac_i$ and each $j \in \Nc_i$, if $k \in \Kc_j$, receive $ \lsem  (z_j^{\tau+1})_{k_j} \rsem_{0}$ from agent $j$.
    \item Compute $\lsem  (\zeta^{\tau+1})_{\Ac_i}  \rsem_{0}$ according to~\eqref{eq:encryptedZetaUpdate}.
    \item For each $j\in \Nc_i$, send $\lsem  (\zeta^{\tau+1})_{\Ac_i\cap\Kc_j}  \rsem_{0}$ to agent~$j$ secured by AES.
    \item For each $j\in \Nc_i$, receive $\lsem (\zeta^{\tau+1})_{\Ac_j \cap \Kc_i} \rsem_{0}$ from agent~$j$.
    \item Compute $\lsem \lambda_i^{\tau+1} \rsem_{0}$ according to~\eqref{eq:encrpytedLambdaUpdate}.
\end{enumerate}
\end{alg}

An encrypted version of the initialization routine in Algorithm~\ref{alg:plainInit} can be obtained analogously:

\begin{alg}%
\label{alg:encryptedInit}
Encrypted ADMM initialization at agent $i$.\!
    \begin{enumerate}
    \item Set $\cipher{\lambda_i^0}_{0}:=\cipher{0}_{0}$ and, for each $k \in \Ac_i$, encrypt $(\hat{\alpha}_i)_{k_i}$ via $\Enc_{0}$ and set $\cipher{(\zeta^0)_{k}}_{0}:=\cipher{(\hat{\alpha}_i)_{k_i}}_{0}$.
    \item For each $k \in \Ac_i$ and each $j \in \Nc_i$, if $k \in \Kc_j\setminus \Ac_j$,  send $\cipher{(\hat{\alpha}_i)_{k_i}}_{0}$ to agent~$j$ secured by AES.
    \item For each $k \in \Kc_i \setminus \Ac_i$, receive $\cipher{(\hat{\alpha}_j)_{k_j}}_{0}$ from the agent $j$ for which $k\in \Ac_j$, and set $\cipher{(\zeta^0)_{k}}_{0}:=\cipher{(\hat{\alpha}_j)_{k_j}}_{0}$.
\end{enumerate}
\end{alg}

Now, the encrypted ADMM scheme requires one crucial additional element, which is not necessary in the plaintext version.
In fact, as indicated above, agent $i$ shall get access to $\alpha_i^\ell$ after evaluating a user-defined number of $\ell$ ADMM iterations (via Alg.~\ref{alg:encryptedADMM}). Hence, a final key switch $\cipher{\alpha^\ell_i}_{0\rightarrow i}$ is required once. Clearly, this switch must not be performed by agent $i$ since it requires $\Enc_{i}(\sk_{0})$ and since agent $i$ holds $\sk_{i}$. Thus, given $\Enc_{i}(\sk_{0})$, agent $i$ could apply $\Dec_{i}$ to obtain $\sk_{0}$, which would render the encrypted implementation of the algorithms useless.
As a remedy, agent $i$ selects a neighbor $j\in \Nc_i$ and sends $\cipher{\alpha^\ell_i}_{0}$ to it secured by AES.
The neighbor then performs the key switch $\cipher{\alpha^\ell_i}_{0\rightarrow i}$ and returns $\cipher{\alpha^\ell_i}_{i}$ (without the need of an additional encryption) to agent $i$, who is finally able to decrypt (and apply) the result. Remarkably, in the preceding iteration $\tau=\ell-1$, Algorithm~\ref{alg:encryptedADMM} can be terminated after the first step since the results of all upcoming steps are of no avail.

Since multiple keys are involved in the encrypted ADMM implementation, we briefly summarize the key management. Importantly, only the operator holds $\sk_0$. However, since computations are carried out by the agents and since communications are secured by AES, the operator gets no access to corresponding ciphertexts. Now, each agent $i$ holds the secret key $\sk_i$ to its own instance of the cryptosystem, which is required for the final decryption of $\cipher{\alpha^\ell_i}_{i}$. In order to perform the preceding key switch, for each agent $i$, one selected neighbor obtains the switching key in terms of $\Enc_{i}(\sk_{0})$.

\subsection{Security guarantees}

The encrypted ADMM scheme in Section~\ref{subsec:encryptedADMMviaKeySwitches} has been designed based on the security goals formulated in Section~\ref{sec:problemStatement}. However, having specified the encrypted algorithms, we shall briefly confirm the achievement of the goals in light of the considered attacker model (i.e., honest-but-curious).
Clearly, the first goal is achieved since $\beta_i$, $H_i$,~$F_i$, $G_i$, and $E_i$ are only processed by agent $i$ and since only encrypted versions of $\alpha_i^\ell$ are processed at a neighboring agent.
The second goal is satisfied since the operator only provides $\cipher{\delta_i}_0$ to agent $i$. The third goal is achieved since $\gamma_i$ is only processed in an encrypted manner at agent $i$ without ever decrypting it. An analogue argument can be presented with respect to the fourth goal. In fact, while some entries of $\zeta$ are processed by multiple agents, none of them holds the secret key $\sk_0$ required for decryption. Still, we might have the situation that only a single agent $i$ makes use of some $(\zeta)_k$. Due to $|\Ic_k|=1$, we then obtain $(\zeta^{\tau+1})_k=(z_i^{\tau+1})_{k_i}$ according to \eqref{eq:zetaUpdateLocalAvg}. If, in addition, $k\in\Ac_i$, then $(\zeta^{T})_k=(\alpha^\ell)_{k_i}$. Hence, in this special situation, agent $i$ will unavoidably get access to some entries of $\zeta^\ell$, which explains the exception within the goal's definition.

Now, the statements above are only valid if each agent does not get unintended access to secret keys beyond $\sk_{i}$ (and AES keys). In the honest-but-curious setup, the only way to obtain additional knowledge about other keys could result from insecure implementations of the key switches. In fact, it is important to note that key switches can only be securely carried out if so-called key-cycles\footnote{According to \cite{alamati2016three}, a cycle of $k$ encrypted keys is a set of the form $\{\Enc_{1}(\sk_{2}),\Enc_{2}(\sk_{3}),\dots,\Enc_{k-1}(\sk_k),\Enc_{k}(\sk_1)\}$.} are avoided~\cite{alamati2016three}.
In order to perform the final key switch $\cipher{\alpha^\ell_j}_{0 \rightarrow j}$ for one or more of its neighbors, agent $i$ may hold one or more ``switching keys'' from the set  $\{ \Enc_{j}(\sk_{0}) \,|\, j \in \Nc_i \}$.
Clearly, since $\sk_{0}$ is fixed and $j \neq 0$, the set can never contain a key-cycle. However, the switching keys have to be generated and distributed by the operator.
Hence, sending $\Enc_{j}(\sk_{0})$ to some agent $i\neq j$ also requires AES since, otherwise, the agent $j$ might eavesdrop the communication and obtain $\sk_0$. Remarkably, also the operator has no access to a key-cycle since all generated switching keys are from the set $\{ \Enc_{i}(\sk_{0}) \,|\, i \in \Mc \}$.

While the proposed scheme is secure with respect to the security goals and the specified attacker model, some weaknesses arise if the agents slightly deviate from the honest evaluation of the algorithms. Most prominently, agent $i$ could send any other ciphertext of the same dimension as $\cipher{\alpha^\ell_i}_{0}$ to agent $j$ for the final key switch $\cipher{\cdot}_{0\rightarrow i}$ and thus obtain access to any intermediate result of Algorithms~\ref{alg:encryptedADMM} or \ref{alg:encryptedInit}. Fortunately, this vulnerability can be easily avoided by outsourcing the computation of $\cipher{ z_i^{\tau+1} }_{0}$ in Algorithm~\ref{alg:encryptedADMM} to the neighboring agent who, so far, only provided the key switch $\cipher{\alpha_i^\ell}_{0\rightarrow i}$. This way, agent $i$ can no longer present unintended data for the key switch.
However, in order to obey the first security goal, the matrices $\rho \Gamma_i^{11}$, $\Gamma_i^{11}$, and $\Gamma_i^{12} E_{i} -\Gamma_i^{11} F_i$ for the $z_i$ update now need to be encrypted. As a consequence, \eqref{eq:encryptedZiUpdate} gets replaced by
\begin{align*}
   \lsem z_i^{\tau+1} \rsem_{0} &= \left(\lsem \rho \Gamma_i^{11} \rsem_{0} \otimes \lsem (\zeta^\tau)_{\Kc_i} \rsem_{0} \right) \ominus \left(\lsem \Gamma_i^{11} \rsem_{0} \otimes \lsem \lambda_i^\tau \rsem_{0} \right) \\
   &\quad \oplus  \left( \lsem \Gamma_i^{12} E_{i} -\Gamma_i^{11} F_i \rsem_{0} \otimes  \lsem p_i \rsem_{0} \right),
\end{align*}
which now involves encrypted multiplications as in~\eqref{eq:mulHomomorphism}.
In addition, the data exchange is slightly more complex, as data from neighbors of neighbors might be involved. Apart from that, implementing the indicated variant is straightforward.

\section{Case study on robot formation}%
\label{sec:robotFormation}

Next, we illustrate the application of our encrypted ADMM scheme for privacy-preserving robot formation. More precisely, we consider a MAS consisting of $M$ mobile robots, who should drive in a certain formation, specified by the system operator, without actually knowing the formation and without revealing individual positions to other agents or the operator. To prepare the case study, we initially specify the system dynamics as well as the control task and then reformulate the task by means of general consensus. Finally, we give some insights on the encrypted implementation, including performance measures.

\subsection{System dynamics and control task}

For our case study, we consider mobile robots moving on a plane with linear dynamics of the form
\begin{subequations}%
\label{eq:modelRobots}
\begin{align}
    x_i (t+1) &= A_i x_i (t) + B_i u_i (t) \\
    y_i (t) &= C_i x_i (t),
\end{align}
\end{subequations}
where $t\in \N$ denotes a discrete time step.
For simplicity, we assume that all agents have double-integrator dynamics (discretized with a sample time of $\Delta t:=1$) in both longitudinal and
transversal
direction, with $u_i(t) \in \R^2$ reflecting accelerations in the corresponding directions.
Further, the outputs $y_i(t) \in \R^2$ reflect the robot's position in a global coordinate system on the plane. We omit details on the corresponding matrices
$A_i$, $B_i$, and $C_i$ for brevity.
We specify the formation based on the robot's relative positions. More precisely, for neighboring agents, the operator specifies a desired (possibly time-varying) displacement $d_{ij}(t) \in \R^2$ ideally reflecting $y_i(t)-y_j(t)$. We here assume consistency of the desired displacements and, in particular, ${d_{ij}(t)=-d_{ji}(t)}$. Now, in order to efficiently achieve the formation, each agent $i$ is supposed to minimize the cost function
\begin{equation}
\label{eq:costFormationU}
    r\,\|U_i(t)- U_i(t-1)\|_2^2 + \!\sum_{j \in \Nc_i} \! \|Y_i(t)- Y_j(t)-D_{ij}(t)\|_2^2
\end{equation}
on a receding horizon with $N \in \N$ prediction steps, where
$$
U_{i}(t):=\begin{pmatrix}    u_{i}(t) \\
    u_i(t+1) \\
    \vdots\\
    u_{i}(t+N-1)
\end{pmatrix} \quad \text{and} \quad Y_{i}(t):=\begin{pmatrix}
    y_{i}(t+1) \\
    y_i(t+2) \\
    \vdots\\
    y_{i}(t+N)
\end{pmatrix}
$$
denote stacked input and output sequences, respectively, and where the sequence $D_{ij}(t)$ is defined analogously.
A special role is taken by the first agent, who acts as a leader, who is additionally supposed to follow a (possibly time-varying) reference trajectory. Hence, for $i=1$, the term $\eta\,\|Y_1(t)- Y_{1,\text{ref}}(t)\|_2^2$ is added to~\eqref{eq:costFormationU}, where $Y_{1,\text{ref}}(t)$ reflects a sequence of reference outputs $y_{1,\text{ref}}(t)$. The weighting factors $r$ and $\eta$ can be used to emphasize certain terms in the cost functions.

\subsection{Formulation as general consensus}

We next aim for a (re)formulation of the control task in the form~\eqref{eq:generalizedConsensusOP} with the specifications~\eqref{eq:fiQuadratic} and \eqref{eq:zipiStructures}. Clearly, the cost function in~\eqref{eq:costFormationU} (also with the extension for $i=1$) is quadratic, which well harmonizes with the first terms of $f_i$ in~\eqref{eq:fiQuadratic}. Moreover, it is straightforward to condense the linear dynamics~\eqref{eq:modelRobots} into a relation of the form
\begin{equation}
\label{eq:relationYxU}
Y_{i}(t)= \Oc_{i} x_{i}(t)+ \Tc_{i} U_{i}(t),
\end{equation}
which links the input and output sequences via the current state $x_i(t) \in \R^4$. The constraint~\eqref{eq:relationYxU} well aligns with the ``indicator-term'' in~\eqref{eq:fiQuadratic}. Hence, it mainly remains to specify suitable local variables $z_i$ and parameters $p_i$.
Clearly, with~\eqref{eq:modelRobots} in mind, $z_i$ should contain $U_i(t)$, $Y_i(t)$, and $Y_j(t)$ for all neighbors $j \in \Nc_i$. Moreover, the parameter vector $p_i$ should reflect the desired displacements $D_{ij}(t)$ to all neighbors as well the former input $u_i(t-1)$ (entering via $U_i(t-1)$). Taking also~\eqref{eq:relationYxU} into account, $x_i(t)$ has to be added to $p_i$. In summary, we obtain

\begin{equation}
\label{eq:ziPiRobtos}
    z_{i}:=\begin{pmatrix}
    U_{i}(t) \\
     Y_{i}(t) \\
 Y_{j}(t)\,\,\, \forall j\in \Nc_i
\end{pmatrix} \,\,\,\text{and} \,\,\, p_i:=\begin{pmatrix}
u_{i}(t-1)\\
     x_{i}(t)\\
    D_{ij}(t)\,\,\, \forall j \in \Nc_i
\end{pmatrix},\!
\end{equation}
where $p_1$ is augmented by $Y_{1,\text{ref}}(t)$. Now, rewriting~\eqref{eq:costFormationU} and \eqref{eq:relationYxU} in the form of \eqref{eq:fiQuadratic} by selecting appropriate matrices $H_i$, $G_i$, $F_i$, and $E_i$ is straightforward and omitted for brevity.
More importantly, $z_i$ and $p_i$ as in~\eqref{eq:ziPiRobtos} clearly offer the structures \eqref{eq:zipiStructures}, since
\begin{equation}
\nonumber
    \alpha_{i}:=\begin{pmatrix}
    U_{i}(t) \\
     Y_{i}(t)
\end{pmatrix} \quad \text{and} \quad  \beta_i:=\begin{pmatrix}
u_{i}(t-1)\\
     x_{i}(t)
\end{pmatrix}
\end{equation}
reflect individual variables and parameters of agent $i$, respectively.

\subsection{Numerical experiments}
For our numerical experiments, we will consider the three communication graphs in Figure~\ref{fig:threeGraphs} with $M\in\{8,9\}$ agents and also aim for the illustrated (time-invariant) formations. More specifically, the desired formation for the ring graph (Fig.~\ref{fig:threeGraphs}) is a regular octagon with a circumcircle of radius $10$ and the other formations are similar apart from the central leader.
For all three setups, the leader agent is supposed to follow a reference trajectory along the $y_1$-axis. In order to simplify a comparison of the setups, we specify
 $y_{1,\text{ref}}(t):=(10+t \quad 0)^\top$ for the ring graph and  $y_{1,\text{ref}}(t):=(t \quad 0)^\top$ for the two other graphs (to account for the central leader). In addition, we choose a prediction horizon of $N=4$ and the weighting factors $r=0.1$ and $\eta=10$.
 Further, we randomly define the robots' initial positions $y_i(0)$ in the box $[-10,10]^2$ and set the initial velocities (i.e., the remaining entries of $x_i(0)$) as well as $u_i(-1)$ to zero.
 For the ADMM schemes, we set $\rho=0.2$ and evaluate $\ell=5$ iterations (per time step).
 Finally, we initialize the schemes with
$\hat{\alpha}_i:=\begin{pmatrix}
    0 &\dots& 0 & y_i^\top(0) & \dots & y_i^\top(0)
\end{pmatrix}^\top$
at $t=0$ and reuse $\alpha_i^\ell$ from the previous time step as an initialization for all upcoming $\ell$.

\begin{figure}[t!]
    \centering
    \begin{subfigure}[c]{0.5\textwidth}
    \captionsetup{justification=centering}
    \subcaption{Ring graph}%
    \label{subfig:results_ring}
    \vspace{1mm}
    \includegraphics{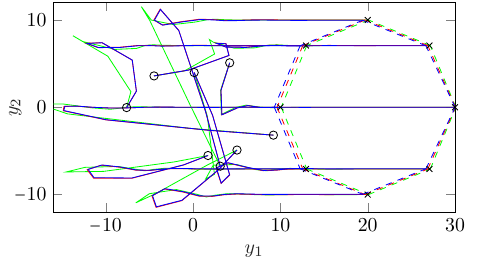}
    \end{subfigure}

    \centering
    \begin{subfigure}[c]{0.5\textwidth}
    \captionsetup{justification=centering}
    \subcaption{Star graph}%
    \label{subfig:results_star}
    \vspace{1mm}
    \includegraphics{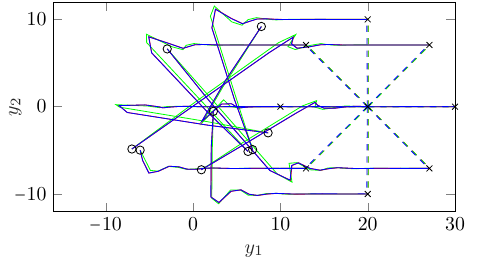}
    \end{subfigure}

    \centering
    \begin{subfigure}[c]{0.5\textwidth}
    \captionsetup{justification=centering}
    \subcaption{Generic graph}%
    \label{subfig:results_generic}
    \vspace{1mm}
    \includegraphics{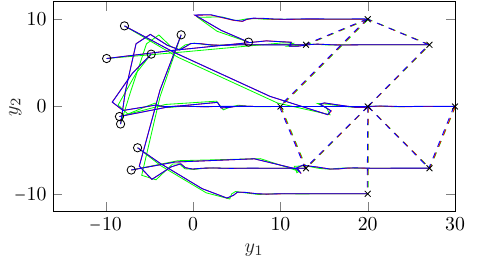}
    \end{subfigure}

    \caption{Experimental results for the graphs from Fig.~\ref{fig:threeGraphs}. The centralized, distributed, and encrypted solution are depicted in green, red, and blue, respectively. Solid lines represent the robot's trajectories, with circles marking the initial position. Crosses mark the ideal position and dashed lines (in green, red, and blue) show the actual formation at $t=20$.}%
    \label{fig:resultsThreeGraphs}
\end{figure}

We are now ready to briefly investigate the plaintext implementation of the ADMM scheme as a reference for the encrypted version studied below.
With the specifications above, we obtain the red trajectories in Figure~\ref{fig:resultsThreeGraphs} for the three graphs. Clearly, the formations are quickly achieved, and the leader follows the reference. Moreover, the trajectories are close to those in green, resulting from a centralized solution of~\eqref{eq:generalizedConsensusOP}. In fact, relatively large deviations only occur for the ring graph, which is no surprise since the connectivity of the graph is low.
Regarding the encrypted implementation, it remains to specify the utilized cryptosystem and the mapping~\eqref{eq:scalingRoundingModq} onto the integer message space.
As indicated above, we consider a leveled FHE scheme based on (ring) LWE. More precisely, we opted for the CKKS cryptosystem~\cite{cheon2017homomorphic} implemented in the OpenFHE library~\cite{OpenFHE}.

While omitting technical details, we will briefly discuss our choice of the modulus $q$ and the dimension $n$ of the secret keys $\sk_i$ because security mainly relies on these parameters.
Now, the choice of $q$ depends on the selected scaling factor $s$ in \eqref{eq:scalingRoundingModq} and the number of encrypted multiplications because they (implicitly) increase the scaling.
In this context, the updates \eqref{eq:encryptedZetaUpdate}, \eqref{eq:encrpytedLambdaUpdate}, and \eqref{eq:encryptedZiUpdate} amount to three successive
(i.e., non-parallelizable) multiplications per ADMM iteration.
One extra multiplication is required during the final key switch after $\ell$ iterations. Thus, our implementation must support $3\ell+1=16$ successive multiplications.
Moreover, in order to counteract noise-flooding in CKKS, an additional ``internal'' scaling $\sigma$ is required.
Thus, the overall scaling becomes $(s\sigma)^{16}$.

Now, under the reasonable assumption that the agents' inputs and outputs remain in the box $[-100,100]^2$,
we can exclude an overflow
if $2\cdot 100 \cdot (s\sigma)^{16} \leq q$, where the factor $2$ reflects the signed inputs and outputs.
In the CKKS implementation in OpenFHE, $q$ has to be provided in the form $q=q_0 \cdot S^L$ by specifying the base modulus $q_0$, the overall scaling factor $S=s\sigma$, and the level $L$, where $q_0>S$ is required. Based on the obvious choice $L=16$, we set $S=2^{23}$ and $q_0=2^{25}$ for our experiments, which retains an accuracy of about three decimal digits (after the separator)
in the output of the algorithm compared to
the plaintext result.
Finally, we consider two choices for the key dimension, namely
$n\in\{2^8,2^{13}\}$, which result in a $40$ and $70$ bit security guarantee, respectively, according to the estimator proposed in \cite{albrecht2015concrete}. Note that our choices reflect a trade-off between performance and security, i.e, a recommended $128$ bit security would lead to (more) impractical encrypted computations. In our case, however, the time frame during which an attempt to break the cipher
makes sense is relatively short. In fact, breaking both $40$ and $70$ bit security requires a significant amount of time even for sophisticated attackers with specialized and powerful computing capabilities (minutes to days in the first case; several years to millennia in the second).

With the parameters above, we obtain the blue trajectories in Figure~\ref{fig:resultsThreeGraphs} when applying our encrypted ADMM scheme to the three examples. It can be seen that the accuracy (mainly depending on $s$) suffices to reproduce the red plaintext trajectories accurately in all cases.
However, the required computational effort can be an issue with respect to real-time requirements.
In fact, even for parallelized computations using eight processors (AMD Ryzen 7 5800H, 8$\times$ \unit[3.2]{GHz}) on a computer with \unit[8]{GB} of RAM, we were able to meet the sampling time of $\Delta t = \unit[1]{s}$ only for $n=2^8$ and the ring graph setting while neglecting communications times (for exchanging ciphertexts of around \unit[0.1]{MB} each). For larger ring dimensions $n$ and more complex communication graphs, the required computation times can increase significantly. For instance, $n=2^{13}$ results in execution times of around $\unit[28]{s}$ already for the ring graph (where now ciphertext of around \unit[0.8]{MB} have to be exchanged).
However, even for the more complex cases, real-time capability might be achievable by, e.g., involving other cryptosystems such as secure multi-party computation or hardware accelerators.

\section{Conclusions and Outlook}%
\label{sec:ConclusionOutlook}

We presented a privacy-preserving cooperative control scheme realized via encrypted ADMM iterations. Due to the distributed nature of cooperative control, such a solution may be valuable in different scenarios. In comparison to existing solutions, our focus is on a generic optimization problem while privacy guarantees are more strict. That is, we require that agents (or the operator) are only allowed to know their individual quantities and no additional information can be extracted from the received data.
To this end, we mainly build on homomorphic encryption, which enables privacy-preserving computations on data exchanged between the agents. In order to manage data access, a special functionality called key switching is used.

We illustrated the effectiveness of our scheme for robot formation control, where only small deviations from a centralized solution arise due to a limited number of ADMM iterations and quantization errors. However, real-time capability is often a challenge due to the computational burden imposed by (current) homomorphic cryptosystems. Therefore, real-world applications are currently restricted to large sample times or agents with significant computational capabilities. Nonetheless, theoretical advances and the use of hardware accelerators may change that \cite{marcolla2022survey}.

Future research is threefold.
First, we plan to apply our scheme to further applications beyond mobile robots.
Second, inequality constraints may be additionally considered during the optimization. Third, a stronger attacker model may be addressed, which considers the corruption of multiple agents or malicious behavior. Different cryptographic techniques like secure multi-party computation will be of interest in this context.

\Paragraph{Funding:}
Financial support by the German Research Foundation (DFG) and the Daimler and Benz Foundation under the grants SCHU 2940/4-1, SCHU 2940/5-1, and 32-08/19 is gratefully acknowledged.

\bibliographystyle{plainnat}

\begin{thebibliography}{28}
\providecommand{\natexlab}[1]{#1}
\providecommand{\url}[1]{\texttt{#1}}
\expandafter\ifx\csname urlstyle\endcsname\relax
  \providecommand{\doi}[1]{doi: #1}\else
  \providecommand{\doi}{doi: \begingroup \urlstyle{rm}\Url}\fi

\bibitem[Al~Badawi et~al.(2022)Al~Badawi, Bates, Bergamaschi, Cousins,
  Erabelli, Genise, Halevi, Hunt, Kim, Lee, Liu, Micciancio, Quah, Polyakov,
  R.V., Rohloff, Saylor, Suponitsky, Triplett, Vaikuntanathan, and
  Zucca]{OpenFHE}
Ahmad Al~Badawi, Jack Bates, Flavio Bergamaschi, David~Bruce Cousins, Saroja
  Erabelli, Nicholas Genise, Shai Halevi, Hamish Hunt, Andrey Kim, Yongwoo Lee,
  Zeyu Liu, Daniele Micciancio, Ian Quah, Yuriy Polyakov, Saraswathy R.V., Kurt
  Rohloff, Jonathan Saylor, Dmitriy Suponitsky, Matthew Triplett, Vinod
  Vaikuntanathan, and Vincent Zucca.
\newblock Open{FHE}: Open-{S}ource {F}ully {H}omomorphic {E}ncryption
  {L}ibrary.
\newblock In \emph{Proceedings of the 10th Workshop on Encrypted Computing \&
  Applied Homomorphic Cryptography}, WAHC'22, page 53–63, New York, NY, USA,
  2022. Association for Computing Machinery.
\newblock ISBN 9781450398770.

\bibitem[Alamati and Peikert(2016)]{alamati2016three}
Navid Alamati and Chris Peikert.
\newblock Three’s compromised too: Circular insecurity for any cycle length
  from ({R}ing\mbox{-}){LWE}.
\newblock In \emph{Advances in Cryptology: 36th Annual International Cryptology
  Conference (CRYPTO), Part II}, pages 659--680. Springer, 2016.

\bibitem[Albrecht et~al.(2015)Albrecht, Player, and
  Scott]{albrecht2015concrete}
Martin~R. Albrecht, Rachel Player, and Sam Scott.
\newblock On the concrete hardness of learning with errors.
\newblock \emph{Journal of Mathematical Cryptology}, 9\penalty0 (3):\penalty0
  169--203, 2015.

\bibitem[Alexandru et~al.(2018)Alexandru, Morari, and
  Pappas]{Alexandru2018_CDC}
Andreea~B. Alexandru, Manfred Morari, and George~J. Pappas.
\newblock Cloud-based {MPC} with encrypted data.
\newblock In \emph{Proc. of the 57th {C}onference on {D}ecision and {C}ontrol},
  pages 5014--5019, 2018.

\bibitem[Alexandru et~al.(2019)Alexandru, Darup, and Pappas]{Alexandru2019_CDC}
Andreea~B. Alexandru, Moritz~Schulze Darup, and George~J. Pappas.
\newblock Encrypted cooperative control revisited.
\newblock In \emph{Proc. of the 58th IEEE Conference on Decision and Control},
  pages 7196--7202, 2019.

\bibitem[Boyd et~al.(2011)Boyd, Parikh, Chu, Peleato, and
  Eckstein]{boyd2011admm}
Stephen Boyd, Neal Parikh, Eric Chu, Borja Peleato, and Jonathan Eckstein.
\newblock Distributed {O}ptimization and {S}tatistical {L}earning via the
  {A}lternating {D}irection {M}ethod of {M}ultipliers.
\newblock \emph{Foundations and Trends{\textregistered} in Machine learning},
  3\penalty0 (1):\penalty0 1--122, 2011.

\bibitem[Cheon et~al.(2017)Cheon, Kim, Kim, and Song]{cheon2017homomorphic}
Jung~Hee Cheon, Andrey Kim, Miran Kim, and Yongsoo Song.
\newblock Homomorphic encryption for arithmetic of approximate numbers.
\newblock In \emph{Advances in Cryptology--ASIACRYPT 2017: 23rd International
  Conference on the Theory and Applications of Cryptology and Information
  Security, Hong Kong, China, December 3-7, 2017, Proceedings, Part I 23},
  pages 409--437. Springer, 2017.

\bibitem[Cheon et~al.(2018)Cheon, Han, Kim, Kim, and
  Song]{cheon2018bootstrapping}
Jung~Hee Cheon, Kyoohyung Han, Andrey Kim, Miran Kim, and Yongsoo Song.
\newblock Bootstrapping for approximate homomorphic encryption.
\newblock In Jesper~Buus Nielsen and Vincent Rijmen, editors, \emph{Advances in
  Cryptology -- EUROCRYPT 2018}, pages 360--384, Cham, 2018. Springer
  International Publishing.
\newblock ISBN 978-3-319-78381-9.

\bibitem[Farokhi et~al.(2017)Farokhi, Shames, and Batterham]{Farokhi2017}
Farhad Farokhi, Iman Shames, and Nathan Batterham.
\newblock Secure and private control using semi-homomorphic encryption.
\newblock \emph{Control Engineering Practice}, 67:\penalty0 13--20, 2017.

\bibitem[Hassan et~al.(2019)Hassan, Rehmani, and Chen]{hassan2019differential}
Muneeb~U. Hassan, Mubashir~Husain Rehmani, and Jinjun Chen.
\newblock Differential privacy techniques for cyber physical systems: a survey.
\newblock \emph{IEEE Communications Surveys \& Tutorials}, 22\penalty0
  (1):\penalty0 746--789, 2019.

\bibitem[Kim et~al.(2021)Kim, Polyakov, and Zucca]{kim2021revisiting}
Andrey Kim, Yuriy Polyakov, and Vincent Zucca.
\newblock Revisiting homomorphic encryption schemes for finite fields.
\newblock In \emph{Advances in Cryptology: 27th Annual International Conference
  on the Theory and Application of Cryptology and Information Security
  (ASIACRYPT), Part III 27}, pages 608--639. Springer, 2021.

\bibitem[Kim et~al.(2023)Kim, Shim, and Han]{Kim2022_TAC}
Junsoo Kim, Hyungbo Shim, and Kyoohyung Han.
\newblock Dynamic controller that operates over homomorphically encrypted data
  for infinite time horizon.
\newblock \emph{IEEE Transactions on Automatic Control}, 68\penalty0
  (2):\penalty0 660--672, 2023.

\bibitem[Kogiso and Fujita(2015)]{Kogiso2015}
Kiminao Kogiso and Takahiro Fujita.
\newblock Cyber-security enhancement of networked control systems using
  homomorphic encryption.
\newblock In \emph{Proc. of the 54th {C}onference on {D}ecision and {C}ontrol},
  pages 6836--6843, 2015.

\bibitem[Lindell(2020)]{lindell2020secure}
Yehuda Lindell.
\newblock Secure {M}ultiparty {C}omputation.
\newblock \emph{Commun. ACM}, 64\penalty0 (1):\penalty0 86–96, dec 2020.
\newblock ISSN 0001-0782.

\bibitem[Maneesha and Swarup(2021)]{maneesha2021powergridsurvey}
Ampolu Maneesha and K.~Shanti Swarup.
\newblock A survey on applications of alternating direction method of
  multipliers in smart power grids.
\newblock \emph{Renewable and Sustainable Energy Reviews}, 152:\penalty0
  111687, 2021.

\bibitem[Marcantoni et~al.(2023)Marcantoni, Jayawardhana, Chaher, and
  Bunte]{Marcantoni2023_LCSS}
Matteo Marcantoni, Bayu Jayawardhana, Mariano~Perez Chaher, and Kerstin Bunte.
\newblock Secure formation control via edge computing enabled by fully
  homomorphic encryption and mixed uniform-logarithmic quantization.
\newblock \emph{IEEE Control Systems Letters}, 7:\penalty0 395--400, 2023.

\bibitem[Marcolla et~al.(2022)Marcolla, Sucasas, Manzano, Bassoli, Fitzek, and
  Aaraj]{marcolla2022survey}
Chiara Marcolla, Victor Sucasas, Marc Manzano, Riccardo Bassoli, Frank~HP.
  Fitzek, and Najwa Aaraj.
\newblock Survey on fully homomorphic encryption, theory, and applications.
\newblock \emph{Proceedings of the IEEE}, 110\penalty0 (10):\penalty0
  1572--1609, 2022.

\bibitem[Nozari et~al.(2016)Nozari, Tallapragada, and
  Cort{\'e}s]{nozari2016differentially}
Erfan Nozari, Pavankumar Tallapragada, and Jorge Cort{\'e}s.
\newblock Differentially private distributed convex optimization via objective
  perturbation.
\newblock In \emph{2016 American control conference (ACC)}, pages 2061--2066.
  IEEE, 2016.

\bibitem[Oh et~al.(2015)Oh, Park, and
  Ahn]{kwang2015surveymultiagentformationcontrol}
Kwang-Kyo Oh, Myoung-Chul Park, and Hyo-Sung Ahn.
\newblock A survey of multi-agent formation control.
\newblock \emph{Automatica}, 53:\penalty0 424--440, 2015.

\bibitem[Paillier(1999)]{Paillier1999}
Pascal Paillier.
\newblock Public-key cryptosystems based on composite degree residuosity
  classes.
\newblock In \emph{Advances in Cryptology - Eurocrypt '99}, volume 1592 of
  \emph{Lecture Notes in Computer Science}, pages 223--238. Springer, 1999.

\bibitem[Regev(2009)]{regev2009lattices}
Oded Regev.
\newblock On {L}attices, {L}earning with {E}rrors, {R}andom {L}inear {C}odes,
  and {C}ryptography.
\newblock \emph{Journal of the ACM (JACM)}, 56\penalty0 (6):\penalty0 1--40,
  2009.

\bibitem[Schulze~Darup et~al.(2018)Schulze~Darup, Redder, Shames, Farokhi, and
  Quevedo]{SchulzeDarup2018_LCSS}
Moritz Schulze~Darup, Adrian Redder, Iman Shames, Farhad Farokhi, and Daniel~E.
  Quevedo.
\newblock Towards encrypted {MPC} for linear constrained systems.
\newblock \emph{IEEE Control Systems Letters}, 2\penalty0 (2):\penalty0
  195--200, 2018.

\bibitem[Schulze~Darup et~al.(2019)Schulze~Darup, Redder, and
  Quevedo]{SchulzeDarup2019_LCSS}
Moritz Schulze~Darup, Adrian Redder, and Daniel~E. Quevedo.
\newblock Encrypted cooperative control based on structured feedback.
\newblock \emph{IEEE Control Systems Letters}, 3\penalty0 (1):\penalty0 37--42,
  2019.

\bibitem[Schulze~Darup et~al.(2021)Schulze~Darup, Alexandru, Quevedo, and
  Pappas]{SchulzeDarup2021_CSM}
Moritz Schulze~Darup, Andreea~B. Alexandru, Daniel~E. Quevedo, and George~J.
  Pappas.
\newblock Encrypted control for networked systems: An illustrative introduction
  and current challenges.
\newblock \emph{IEEE Control Systems Magazine}, 41\penalty0 (3):\penalty0
  58--78, 2021.

\bibitem[Tian et~al.(2022)Tian, Guo, Sun, and Zhou]{Tian2022PPpowersystems}
Nianfeng Tian, Qinglai Guo, Hongbin Sun, and Xin Zhou.
\newblock Fully privacy-preserving distributed optimization in power systems
  based on secret sharing.
\newblock \emph{iEnergy}, 1\penalty0 (3):\penalty0 351--362, 2022.

\bibitem[Tjell and Wisniewski(2019)]{tjell2019PPDistOpt}
Katrine Tjell and Rafael Wisniewski.
\newblock {P}rivacy {P}reservation in {D}istributed {O}ptimization via {D}ual
  {D}ecomposition and {ADMM}.
\newblock In \emph{2019 IEEE 58th Conference on Decision and Control (CDC)},
  pages 7203--7208, 2019.

\bibitem[Van~Parys and Pipeleers(2017)]{van2017distributed}
Ruben Van~Parys and Goele Pipeleers.
\newblock Distributed {MPC} for multi-vehicle systems moving in formation.
\newblock \emph{Robotics and Autonomous Systems}, 97:\penalty0 144--152, 2017.

\bibitem[Zhang et~al.(2019)Zhang, Ahmad, and
  Wang]{Zhang2019admmDecentralizedOpt}
Chunlei Zhang, Muaz Ahmad, and Yongqiang Wang.
\newblock {ADMM} {B}ased {P}rivacy-{P}reserving {D}ecentralized {O}ptimization.
\newblock \emph{IEEE Transactions on Information Forensics and Security},
  14\penalty0 (3):\penalty0 565--580, 2019.

\end{thebibliography}

\end{document}